\documentclass{article}
\usepackage{graphicx}
\usepackage{openbib}
\usepackage{a4wide}
\usepackage{hyperref}
\parskip = \baselineskip

\begin{document}

\begin{titlepage}
{\Large \bf Recurrence plot statistics and the effect of embedding}

T. K. March,$^1$ S. C. Chapman$^1$ and R. O. Dendy$^{2,1}$

$^1$Space and Astrophysics Group, Department of Physics, Warwick University,
Coventry CV4~7AL, United Kingdom

$^2$UKAEA Culham Division, Culham Science Centre, Abingdon, Oxfordshire OX14 3DB, United Kingdom.

{\it published in Physica D 200 pages 171--184}

\begin{abstract}
Recurrence plots provide a graphical representation of the recurrent patterns in a timeseries, the quantification of which is a relatively new field. Here we derive analytical expressions which relate the values of key statistics, notably determinism and entropy of line length distribution, to the correlation sum as a function of embedding dimension. These expressions are obtained by deriving the transformation which generates an embedded recurrence plot from an unembedded plot. A single unembedded recurrence plot thus provides the statistics of all possible embedded recurrence plots. If the correlation sum scales exponentially with embedding dimension, we show that these statistics are determined entirely by the exponent of the exponential. This explains the results of Iwanski and Bradley (Chaos {\bf8} [1998] 861-871) who found that certain recurrence plot statistics are apparently invariant to embedding dimension for certain low-dimensional systems. We also examine the relationship between the mutual information content of two timeseries and the common recurrent structure seen in their recurrence plots. This allows time-localized contributions to mutual information to be visualized. This technique is demonstrated using geomagnetic index data; we show that the AU and AL geomagnetic indices share half their information, and find the timescale on which mutual features appear.

\end{abstract}
\end{titlepage}

\section{Introduction}

Patterns are ubiquitous in nature, where their presence may imply inherent predictability. As a result there is great interest in developing methods for detecting and quantifying patterns, leading to quantitative measures of structure, similarity, information content, and predictability. Here we consider recurrence plots, which offer a means to quantify the pattern within a timeseries, and also the pattern shared between two timeseries.

Recurrence plots are a method for visualizing recurrent patterns within a timeseries or sequence. They were first proposed in 1981 by Maizel and Lenk \cite{Maizel81} as a method of visualizing patterns in sequences of genetic nucleotides. They have since been introduced into the study of dynamical systems \cite{Eckmann87}, where much effort has been put into building quantification schemes for the plots and for the patterns within them. There are now many quantitative recurrence plot measures available \cite{Webber94,Faure98}. These have been applied with success to patterns as diverse as music \cite{Foote01}, climate variation \cite{Marwan02b}, heart rate variability \cite{Marwan02c}, webpage usage \cite{Bernstein91}, video recognition \cite{Cutler00}, and the patterns in written text and computer code \cite{Church93}.

In outline, a data series $S$ can be considered as a set of $n$ scalar measurements 
\begin{equation}
    S = \{s_1,s_2,s_3,...,s_n\}
\end{equation}
from which a sequence of $N$ $d$-dimensional vectors ${\mathbf{a}_k}$ can be constructed using a procedure known as time-delay embedding. The vectors are defined as
\begin{equation}\label{eq:embed}
    \mathbf{a}_k = \{ s_k, s_{k+\tau},s_{k+2 \tau},...,s_{k+(d-1) \tau}\}
\end{equation}
where $\tau$ is a delay parameter and $d$ is known as the embedding dimension, \cite{Takens81}; these parameters are typically chosen independently of the recurrence plot technique, for example see \cite{Fraser86}. A recurrence plot is constructed by considering whether a given pair of these coordinates are nearby in the embedding space. Typically, the maximum norm is used,
\begin{equation}\label{eq:norm}
    \|\mathbf{a}_i-\mathbf{a}_j\| \equiv \max_k\{ |s_{i+k} - s_{j+k}| \}
\end{equation}
so that the distance between two coordinates equals the maximum distance in any dimension. A recurrence plot is represented by a tensor $T_{ij}^d$ whose elements correspond to the distance between each of the $N^2$ possible pairs of coordinates $\mathbf{a}_i$, $\mathbf{a}_j$ \cite{Eckmann87}:
\begin{equation}\label{eq:trplot}
    T_{ij}^A = \Theta \left( \epsilon - \|\mathbf{a}_i-\mathbf{a}_j\|\right)
\end{equation}
where $\Theta$ is a step function ($0$ for negative arguments, $1$ for positive
arguments). For each pair of coordinates in the series whose separation is less than the threshold parameter $\epsilon$, $T_{ij}$ takes the value unity, which can be plotted as a black dot on an otherwise white graph.

A recurrence plot of independent and identically distributed (IID) data appears as a random scattering of black dots, while a regularly repeating signal (such as a sine wave, e.g. see Fig. 1 of \cite{Iwanski98}) appears as a series of equally spaced, $45^{\circ}$ diagonal black lines. An irregularly repeating signal (such as the output of a chaotic system) typically appears as a pattern of small diagonal lines of varying length. Paling of the plot away from the main diagonal indicates that the longer one observes no repeat of a particular feature, the less likely a repeat is to occur. In this case it follows that probability depends on time, and therefore that the process which generated such data is non-stationary.

In this paper we investigate the statistics of recurrence plots, and their meaning in relation to well understood statistics from nonlinear timeseries analysis. First, we examine the meaning of two of the key statistics in recurrence quantification analysis (RQA), namely the determinism and the entropy of line length distribution \cite{Webber94}, and the effect on them of the time-delay embedding procedure \cite{KantzSchreiber,Takens81}. Iwanski and Bradley \cite{Iwanski98} found that the appearance and statistics of recurrence plots for certain low-dimensional systems are not significantly altered by a small change in the embedding dimension $d$, suggesting that these statistics may be important new invariant characteristics of a system. However, unlike traditional measures where invariance relies on the embedding dimension being sufficiently high, Iwanski and Bradley found the same statistics for an unembedded recurrence plot as for an embedded version. This was further examined by Gao and Cai \cite{Gao00}, who suggested that many recurrence plot statistics may rely on information from a higher embedding dimension than was used to construct the recurrence plot. However this does not completely explain why these quantities appear to be invariant with respect to the embedding dimension; nor whether these quantities are independent of each other, or of other better known measures. This is important, since independent quantities potentially yield new information about a system. In section \ref{sec:transf} we show that all embedded recurrence plots are present within the unembedded plot, accessible via a simple transformation. Using this transformation, we derive in section \ref{sec:stat} the effect of embedding on two RQA statistics: determinism, and entropy of line length distribution. For the case of exponential scaling of the correlation sum [see Eq.(\ref{eq:recurrencerate}) below] with embedding dimension, which might be expected for certain low-dimensional systems, we derive expressions which relate these quantities to the Kolmogorov entropy rate \cite{KantzSchreiber}. This is important for two reasons. First, it provides a new perspective on the physical meaning of these quantities. Second, it can be used to establish baseline values for independent and identically distributed (IID) processes, above or below which a measurement can be said to be significant.

In section \ref{sec:mutual}, we examine the converse question of how well-known statistics from nonlinear timeseries analyis relate to recurrence plots. We demonstrate that a standard algorithm for computing the mutual information between two timeseries is related to counting the number of black dots common to the recurrence plots of the two timeseries in question. This suggests the definition of a new form of cross recurrence plot which, when drawn, allows contributions to the mutual information to be visualized. We apply this technique to a physical system in which issues of predictability and correlation are of practical interest. Earth's geomagnetic activity is monitored by a non-uniformly distributed circumpolar ring of magnetometers, which measure fluctuations in horizontal magnetic field strength due to enhancements in auroral activity. These measurements are compiled to form the AE geomagnetic indices \cite{AE}, of which we consider AU (a proxy for the maximum eastward flowing polar current) and AL (a proxy for the maximum westward flowing current). In common with many other ``real world" timeseries, these timeseries show both low and high dimensional behavior, in this case well defined features on timescales of days (storms) which are embedded in colored noise \cite{Hnat02}.

\section{Effect of Embedding Dimension}
\label{sec:transf}

\begin{figure}
    \centering
    \includegraphics[width=0.3\textwidth]{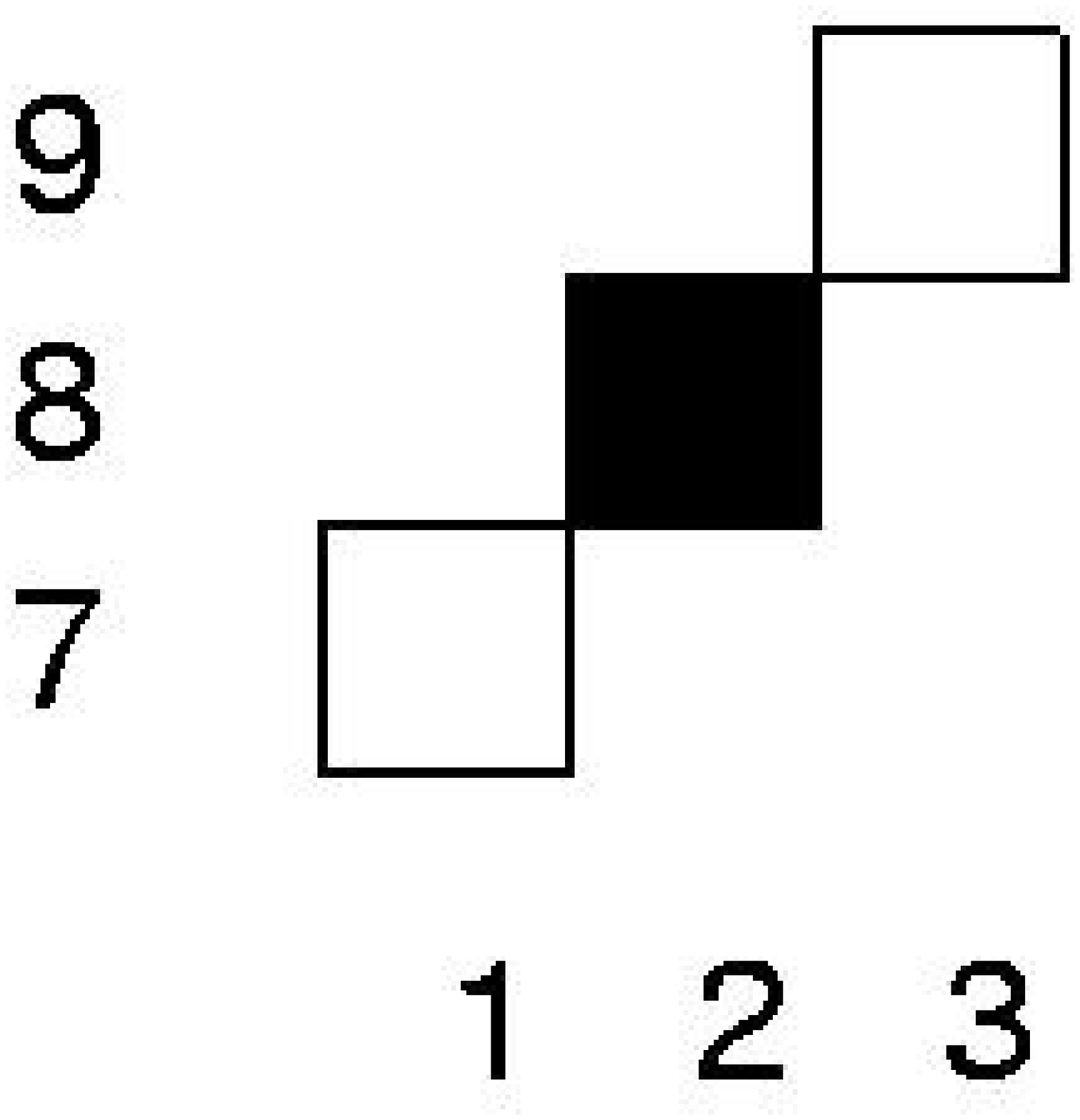}
    \includegraphics[width=0.3\textwidth]{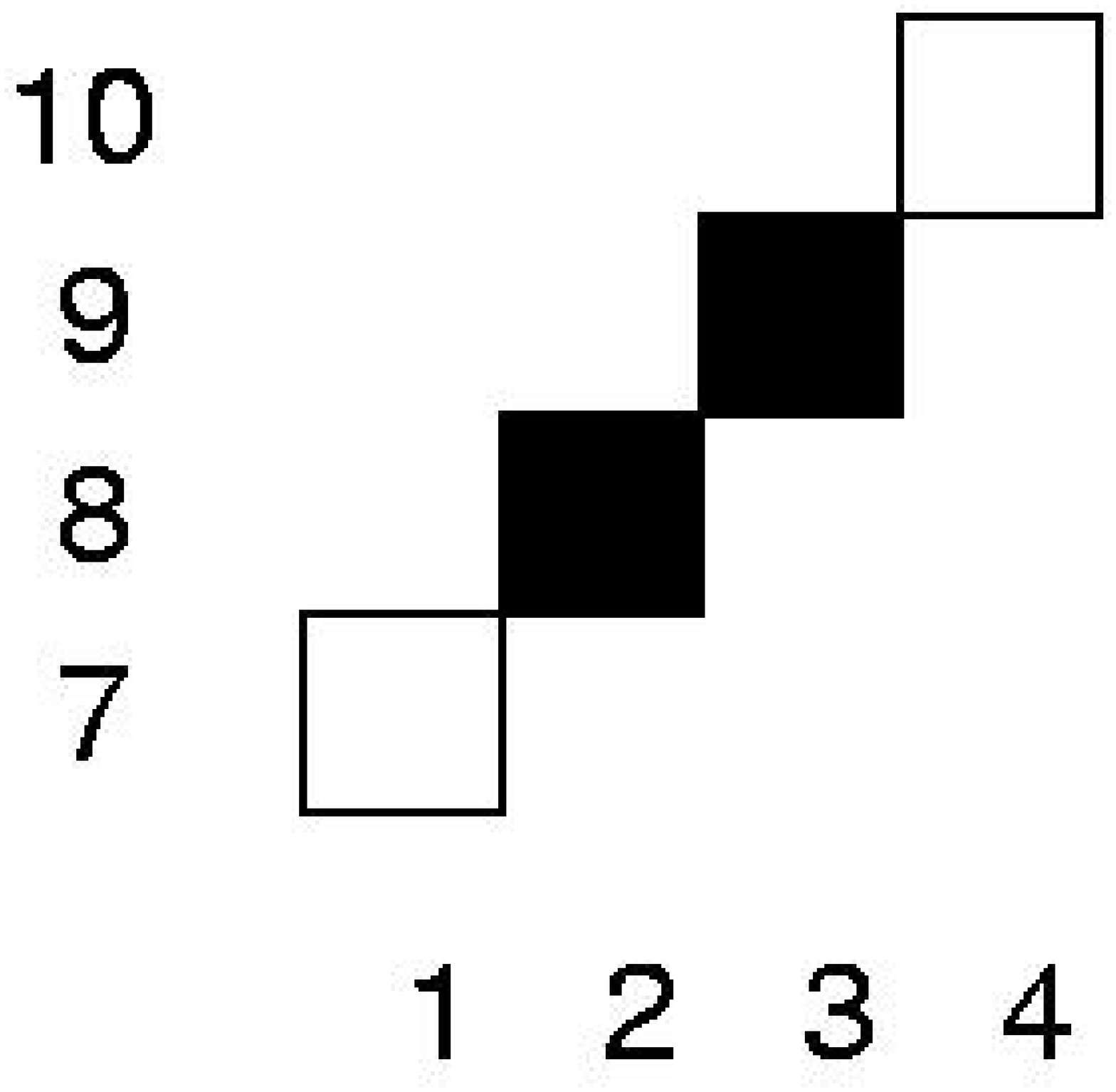}
  \caption{Representation of diagonal lines of length one and two on a recurrence plot, corresponding to pairs of points in the original timeseries.}
  \label{fig:lines}
\end{figure}

We now derive a transformation which generates an embedded recurrence plot from an unembedded recurrence plot. This result is central to the subsequent discussion of the effect of embedding on statistics derived from recurrence plots. A single recurrence on an umbedded $d=1$ plot is represented by a single black dot, corresponding to a pair of data points closer together than $\epsilon$. If we consider Fig. \ref{fig:lines} (left) to represent part of a $d=1$ recurrence plot, the example illustrated relates to points numbered 2 and 8, i.e.
\begin{equation}
	| a_2 - a_8 | < \epsilon
\end{equation}
Figure \ref{fig:lines} (right) shows a line of length two. Still taking $d=1$, the situation represented is
\begin{equation}\label{eq:line2}
	| a_2 - a_8 | < \epsilon \textrm{ and } | a_3 - a_9 | < \epsilon	
\end{equation}
Consider forming coordinates in a $d=2$, $\tau=1$ embedding space [see Eq.~(\ref{eq:embed})]. If we consider Fig. \ref{fig:lines} (left) to represent a region of a $d=2$ recurrence plot, the black dot now represents
\begin{equation}
	\| \mathbf{a}_2 - \mathbf{a}_8 \| < \epsilon
\end{equation}
where $\mathbf{a}_n$ now denotes $\{a_n,a_{n+1}\}$. Using the maximum norm, Eq.~(\ref{eq:norm}), this is equivalent to Eq.~(\ref{eq:line2}). Therefore a single dot in $d=2$ represents a line of length two in $d=1$. The transformation from $d=1$ to $d=2$ thus reduces the length of all diagonal lines by one dot. An isolated dot is removed entirely. 

Formally, we represent the transformation to arbitrary dimension as
\begin{equation}\label{eq:trans1}
    T_{ij}(d) = T_{ij}(1) \times    T_{i+\tau,j+\tau}(1) \times T_{i+2\tau,j+2\tau}(1) \times ... \times T_{i+(d-1)\tau,j+(d-1)\tau}(1)
\end{equation}
An element on the recurrence plot with embedding dimension $d$ is thus related to a diagonal sequence of $d$ elements on the unembedded recurrence plot $T_{ij}(1)$. This transformation enables the conversion of an unembedded recurrence plot into any embedded recurrence plot with any values of $d$ or $\tau$. This suggests that embedding in the construction of recurrence plots is not strictly necessary, since all of the information is contained within the unembedded plot $T_{ij}(1)$; let us refer to this as the parent plot. Rather than performing embedding, information can be extracted directly from this parent plot. Understanding how the information is contained in the parent plot assists in consideration of how various recurrence plot statistics are affected by embedding.

\section{Meaning of Recurrence Plot Statistics}
\label{sec:stat}

Given the transformation derived above, we now consider two of the key statistics of recurrence plots, namely the determinism and the entropy of the diagonal line length distribution. We show that both these statistics are related to the correlation sum, and also relate them to the probability distribution of line lengths on an unembedded plot. In the case of exponential scaling of the correlation sum with embedding dimension, we show that they do not depend on the embedding dimension $d$.

Recurrence quantification analysis (RQA) provides a set of statistical measures which have been proposed to quantify patterns based on the lines and dots visible on a recurrence plot \cite{Webber94}. The fraction of the plot colored black is the most fundamental statistic associated with recurrence plots. This is known as the recurrence rate in RQA, and is known elsewhere as the correlation sum $C_d(\epsilon)$ \cite{KantzSchreiber,Casdagli97}. $C_d(\epsilon)$ is the fraction of pairs of coordinates closer together than $\epsilon$, and is defined by
\begin{equation}\label{eq:recurrencerate}
    C_d^A(\epsilon) = \frac{2}{N(N-1)}\sum_{i=1}^N \sum_{j=i+1}^N \Theta \left( \epsilon - \| \mathbf{a}_i - \mathbf{a}_j\|\right)
\end{equation}
A recurrence plot can be considered to be a two-dimensional pictorial representation of the points that contribute to Eq.~(\ref{eq:recurrencerate}) for a particular value of $\epsilon$.

The remaining statistics in RQA are the fraction of black dots involved in diagonal lines, known as the determinism $D_d$, the entropy of the diagonal line length distribution $E_d$, the ratio of determinism to correlation sum, and the slope of the line of best fit on a graph of recurrence probability versus distance from main diagonal, known as the trend \cite{Webber94}. Except for the trend, these statistics can be related to the probability distribution of diagonal line lengths $P_d(L)$, which is the probability of observing a diagonal black line of length $L$ beginning from a randomly selected element of the recurrence plot. From Eq.~(\ref{eq:trans1}), the distribution of line lengths on an embedded recurrence plot is related to the distribution on an unembedded plot by
\begin{equation}\label{eq:lineprob}
	P_d(L) = P_1(L+d-1)
\end{equation}
Hence any statistic formed from the embedded $P_d(L)$ can be constructed from the unembedded $P_1(L+d-1)$. For example, using Eq.~(\ref{eq:lineprob}), the correlation sum can be written as
\begin{equation}
	C_d = \sum_{L=d}^{\infty}(L-d+1)P_1(L)
\end{equation}
This relationship can be reversed to give
\begin{equation}\label{eq:CofP}
	P_1(L) = C_{L+2}- 2 C_{L+1} + C_{L}
\end{equation}
Hence any statistics derived from $P_1(L)$ can also be derived from the correlation sum, as we now explicitly show.

First we consider the determinism $D_d$ \cite{Webber94}, which was observed to be invariant to embedding dimension by Iwanski and Bradley \cite{Iwanski98}. This is the ratio of black dots included in lines of length greater than unity to the total number of black dots. The determinism $D_d$ quantifies the prevalence of lines, and is believed \cite{Webber94} to quantify how deterministic a system is. This can be related to the probability $C_d$ of observing a black dot in a randomly selected location, and to the probability of observing an isolated black dot. The number of black dots included in lines is equal to the total number of black dots minus the number of isolated black dots (lines of length unity), so we can write
\begin{equation}\label{eq:determinism}
	D_d = \frac{C_d - P_1(d)}{C_d}
\end{equation}
Using Eq.~(\ref{eq:CofP}) to express $P_1(d)$ in Eq.~(\ref{eq:determinism}), we have
\begin{equation}\label{eq:detcool}
	D_d = \frac{2C_{d+1} - C_{d+2}}{C_d}
\end{equation}
Thus the determinism at embedding dimension $d$ can be inferred from knowledge of the correlation sum at nearby embedding dimensions $d$, $d+1$ and $d+2$.

The next statistic in the RQA is the Shannon entropy of the line length distribution \cite{Webber94}. This is defined as
\begin{equation}\label{eq:lineent}
	E_d = - \sum_{L=1}^{\infty} Q_d(L) \ln Q_d(L)
\end{equation}
where $Q_d(L)$ is the probability of observing a line of length $L$ given the fact that a line is observed. This can be related to the probability $P_d(L)$ of observing a line of length $L$, and the probability of observing a line of arbitrary length. Using Eqs.~(\ref{eq:lineprob}) and  (\ref{eq:CofP}) we obtain
\begin{equation}\label{eq:entcorrsum}
	Q_d(L) = \frac{C_{L+d+1}- 2 C_{L+d} + C_{L+d-1}}{C_{d} - C_{d+1}}
\end{equation}
Hence, like the determinism, the Shannon entropy of line length distribution can be obtained from the correlation sum.

\subsection{Exponential scaling of correlation sum}

Suppose we assume that the correlation sum $C_d$ can be expressed as an inverse exponential function of $d$ with exponent $K_2$. This is strictly true for data derived from an IID process, and is observed for many low-dimensional chaotic processes under certain conditions \cite{KantzSchreiber}; in this case $K_2$ is known as the Kolmogorov entropy rate. It has been previously shown that this can be extracted from the distribution of recurrence plot diagonal line lengths $P_d(L)$ \cite{Faure98}. We write the correlation sum as
\begin{equation}\label{eq:scale2}
	C_d = A e^{-K_2 d}
\end{equation}
where we have absorbed the dependence of $C_d$ on the threshold parameter $\epsilon$ into the constant $A$. Substitution of Eq.~(\ref{eq:scale2}) into Eq.~(\ref{eq:CofP}) yields
\begin{equation}\label{eq:eq30}
	P_1(L) = A  (1-e^{-K_2})^2 e^{-K_2 L}
\end{equation}
This implies that $P_1(L)$ is an exponential function of $L$ with the same exponent $K_2$ that governs the dependence of $C_d$ on $d$. This result has been derived independently by an alternative route which considers the divergence of trajectories directly \cite{Gao00}. 

From Eqs.~(\ref{eq:determinism}) and (\ref{eq:eq30}), the determinism $D_d$ can be written
\begin{equation}
	D_d = 1 - \frac{A e^{-K_2 d} (1-e^{-K_2})^2}{A e^{-K_2 d}}
\end{equation}
This simplifies to give
\begin{equation}\label{eq:detsol}
	D_d = 1 - \gamma^2
\end{equation}
where we define $\gamma = (1-e^{-K_2})$. For exponential scaling of $C_d$, the determinism is a constant independent of the embedding dimension $d$ chosen, and is determined by the exponential scaling exponent. Where the correlation sum only exhibits exponential scaling over a limited range of embedding dimensions (such as might be expected for a low-dimensional chaotic process), this expression remains true, since Eq.~(\ref{eq:detcool}) only relies on knowledge of adjacent (in $d$) correlation sums.

To derive the Shannon entropy of line length distribution, Eq.~(\ref{eq:lineent}), we insert Eq.~(\ref{eq:scale2}) into Eq.~(\ref{eq:entcorrsum}) to give
\begin{equation}\label{eq:qdofl}
	Q_d(L) = (1-e^{-K_2}) e^{-K_2 (L-1)}
\end{equation}
which when inserted into Eq.~(\ref{eq:lineent}) gives
\begin{equation}\label{eq:entdef}
	E_d = K_2 \left(\frac{1}{\gamma}-1\right) - \ln \gamma
\end{equation}
As with $D_d$, this is independent of the embedding dimension $d$. However, unlike Eq.~(\ref{eq:detsol}) this expression is only true in the case of perfect exponential scaling.

\begin{figure}
    \centering
    \includegraphics[width=0.6\textwidth]{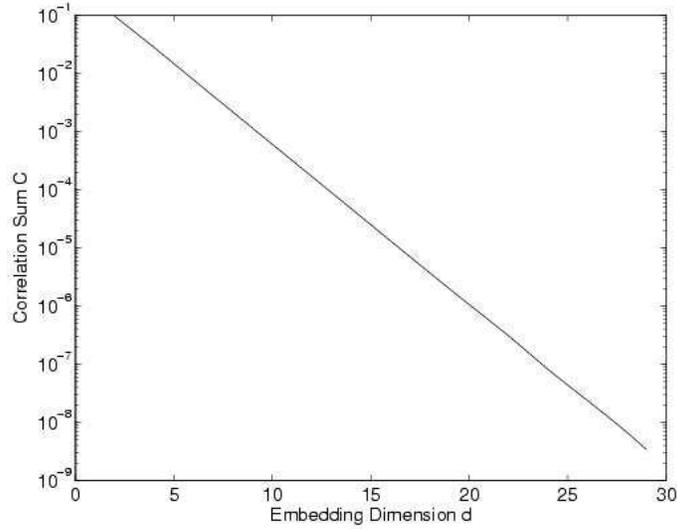}
  \caption{Correlation sum $C_d$ computed as a function of embedding dimension $d$ for $10^5$ samples of the logistic map with $\epsilon = 0.1$. Applying Eq.~(\ref{eq:scale2}) to the measured straight line slope gives $K_2=0.6349 \pm 0.0004$.}
  \label{fig:logiscsum}
\end{figure}

\begin{figure}
    \centering
    \includegraphics[width=0.6\textwidth]{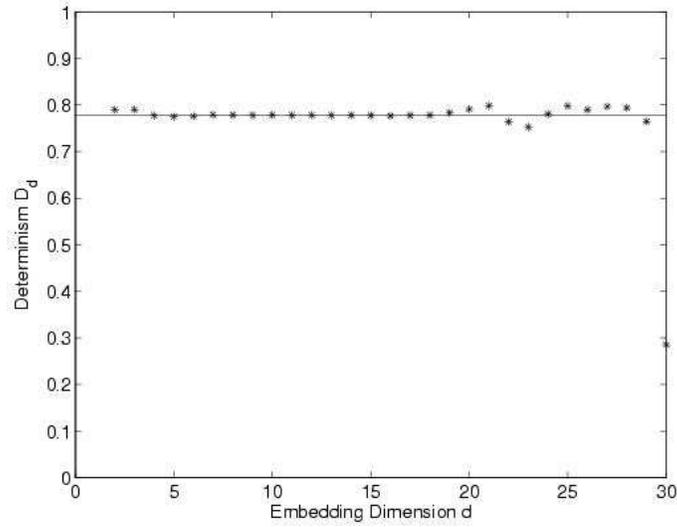}
  \caption{Determinism $D_d$ computed as a function of embedding dimension $d$ for $10^5$ samples of the logistic map with $\epsilon = 0.1$, shown as asterisks. Solid line shows theoretical prediction of 0.7791 obtained from Eq.~(\ref{eq:detsol}) using the measured value of $K_2$ from Fig.~\ref{fig:logiscsum}.}
  \label{fig:logisdet}
\end{figure}

\begin{figure}
    \centering
    \includegraphics[width=0.6\textwidth]{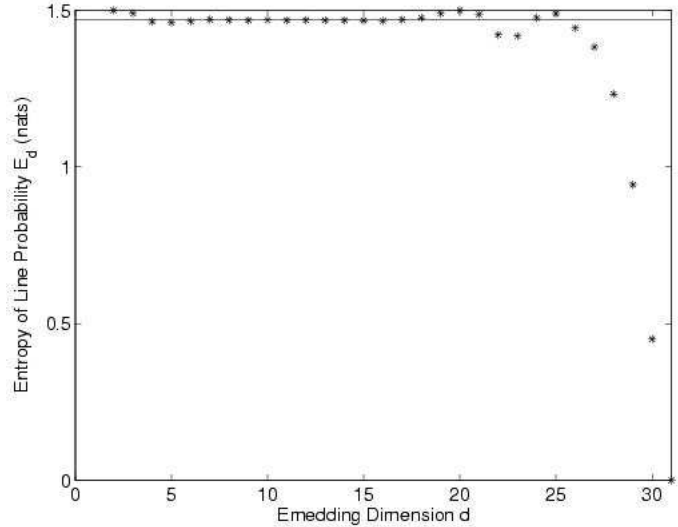}
  \caption{Shannon entropy of line probability distribution $E_d$ computed as a function of embedding dimension $d$ for $10^5$ samples of the logistic map with $\epsilon = 0.1$, shown as asterisks. Solid line shows theoretical prediction of 1.4709 obtained from Eq.~(\ref{eq:entdef}) using the measured value of $K_2$ from Fig.~\ref{fig:logiscsum}.}
  \label{fig:logisent}
\end{figure}

As a demonstration of these results, Fig. \ref{fig:logiscsum} shows the correlation sum computed as a function of embedding dimension for the logistic map, $x_{t+1} = \mu x_t (1-x_t)$, in the chaotic regime with $\mu=4$. This shows reasonable scaling of the correlation sum with dimension, as in Eq.~(\ref{eq:scale2}), and yields $K_2 = 0.6349 \pm 0.0004$. By Eq.~(\ref{eq:detsol}), this implies a value for the determinism $D_d$ of $0.7791 \pm 0.0002$ and by Eq.~(\ref{eq:entdef}) a value for $E_d$ of $1.4709 \pm 0.0006$. These values are shown on Figs.~\ref{fig:logisdet} and \ref{fig:logisent} as the solid lines, while the actual values computed from recurrence plots of the data are shown as asterisks. Until statistical noise becomes important (around $d=25$-$30$), the points lie convincingly on the lines.

An initial exponential distribution of diagonal line lengths remains exponential after embedding, explaining the apparent invariance with respect to $d$ of these statistics for low-dimensional chaotic systems \cite{Iwanski98}. The determinism $D_d$ and the entropy $E_d$ are in this case governed by the exponential scaling exponent of the correlation sum, $K_2$.

A corollary is provided by the results of Zbilut et al. \cite{Zbilut00}, who applied the techniques of recurrence quantification analysis to short sequences of random integers, as well as to the logistic map. There were three sequences considered: (a) consecutive digits of $\pi$; (b) pseudo-random integers generated with MATLAB; (c) experimentally derived random integers, produced by tuning a radio antenna to an empty part of the spectrum \cite{randomorg}. All three were considered with sequence lengths of $N=1000$, $3000$ and $5000$, and only exact matches were considered to constitute recurrences. This corresponds to $\epsilon=0$ in Eq.~(\ref{eq:trplot}), which is only possible when working with integer sequences; for real-valued sequences, $\epsilon$ is limited by numerical precision. It was found that for (a) and (b) the determinism was slightly below $20\%$, and was defined up to $d_0=4$ for $N=1000$, $d_0=5$ for $N=3000$ and $d_0=6$ for $N=5000$. However, (c) had a determinism only slightly above $0\%$, which was defined only up to $d_0=2$ regardless of $N$. The authors suggested that this was possibly due to some innate randomness that sequence (c) possessed, and suggested the RQA as a test to distinguish between physical and pseudo random numbers.

For data drawn from an IID process, the probability of a particular dot being black on the $d=1$ plot is a constant $C_1$, the correlation sum. Referring back to the definition of Eq.~(\ref{eq:scale2}), we can write
\begin{equation}
	\gamma = 1-C_1
\end{equation}
\begin{equation}\label{eq:det}
	D_d = C_1(2-C_1)
\end{equation}
\begin{equation}\label{eq:ent}
	E_d = \frac{1}{C_1}\ln C_1 - \ln (1-C_1)
\end{equation}
These quantities are finite for IID data because a small number of lines are created by chance. To conclude that observed data are non-random, the values measured must be compared with Eqs.~(\ref{eq:det}) and (\ref{eq:ent}) to establish the significance of the result.

Integer sequences can be represented as a string of symbols from an alphabet of size $m$. The probability of observing a black dot in a randomly selected location on the unembedded $d=1$ plot is given by
\begin{equation}
	C_1 = \sum_{i=1}^{m} p_i^2
\end{equation}
where $p_i$ is the probability of observing symbol $i$ from the alphabet. The sequences (a), (b) and (c) were all uniformly distributed so we can write $p_i$ as
\begin{equation}
	p_i = \frac{1}{m}
\end{equation}
and
\begin{equation}
	C_1 = \sum_{i=1}^{m} \frac{1}{m^2} = \frac{1}{m}
\end{equation}

The measured determinism values \cite{Zbilut00} died out above a particular value of $d$, when no diagonal black lines were seen on a finite recurrence plot. To estimate the embedding dimension at which this should occur, we examine the expected number of lines $\left<n\right>$ on an embedded recurrence plot. This is given by the total number of elements on the plot multiplied by the probability of observing, in a randomly selected location, one white dot diagonally followed by $d+1$ black dots on the unembedded recurrence plot:
\begin{equation}
	\left<n\right> = \frac{1}{2} N (N-1) (1-C_1)C_1^{d+1}
\end{equation}
Setting $\left<n\right>$ equal to unity gives an estimate for $d_0$, the dimension where the determinism should die out:
\begin{equation}\label{eq:d}
	d_0 \approx \frac{\log 2 -\log N(N-1) - \log C_1(1-C_1)}{\log C_1}
\end{equation}

Using Eqs.~(\ref{eq:det}) and (\ref{eq:d}) with $m=10$ symbols we obtain $C_1 = 0.1$ and $D_d=19\%$. From Eq.~(\ref{eq:d}), this should be measurable a priori up to $d_0 \approx 4.6$ for $N=1000$, $d_0 \approx 5.6$ for $N=3000$ and $d_0 \approx 6.1$ for $N=5000$, see Fig. \ref{fig:table1}. Comparing these values with the measured results \cite{Zbilut00}, we infer that (a) and (b) behave exactly as would be expected for an IID process with no additional distinguishing properties.

\begin{figure}
\begin{center}
\begin{tabular}[]{|c|r|r|r|r|}
\hline
& \multicolumn{2}{c|}{$m=10$} & \multicolumn{2}{|c|}{$m=100$} \\
\hline
$N$ &Observed $d_0$&Predicted $d_0$&Observed $d_0$&Predicted $d_0$\\
\hline
1000&4&4.6&2&1.8\\
3000&5&5.6&2&2.3\\
5000&6&6.1&2&2.5\\
\hline
\end{tabular}
\caption{Observed and predicted [from Eq.~(\ref{eq:d})] embedding dimension $d_0$ at which determinism $D_d$ drops to zero, as a result of finite sample size $N$, for sequences of symbols from an alphabet of size $m=10$ symbols (columns 2 and 3) and $m=100$ symbols (columns 4 and 5). Observed values from \cite{Zbilut00}: for $m=10$, consecutive digits of $\pi$ and pseudo-random integers generated with MATLAB; for $m=100$, experimentally derived random integer sequence from www.random.org.}
\label{fig:table1}
\end{center}
\end{figure}

To explain the results for the experimentally derived random integers (c), we consider sequences of random integers from the same source \cite{randomorg}. The sequences supplied default to the range 1 to 100, an alphabet of $m=100$ symbols. For this value of $m$ we obtain $C_1=0.01$ and from Eq.~(\ref{eq:det}) we predict a value of determinism $D_d=1.99\%$. This should persist up to $d_0 \approx 1.8$ for $N=1000$, $d_0 \approx 2.3$ for $N=3000$ and $d_0 \approx 2.5$ for $N=5000$; this information is summarized in Fig. \ref{fig:table1}. This agrees with the result reported in \cite{Zbilut00}, so that there is no reason to infer any additional randomness property for (c); the results of recurrence quantification analysis can be explained as a consequence of the different number of symbols in the sequence.

\section{Mutual Information}
\label{sec:mutual}

A recurrence plot can be considered as a visualization of the double summation in the definition of the correlation sum, Eq.~(\ref{eq:recurrencerate}). It is therefore reasonable to expect that a proportion of the statistics derived from recurrence plots would be related to $C_d$. Conversely, it would also be reasonable to expect that existing statistics related to $C_d$ could be derivable from recurrence plots. A recurrence plot would then provide a visualization of any such statistic. As an example we consider the mutual information, which is a nonlinear measure of correlation between two (or more) discrete timeseries. The mutual information $I^{AB}$ between timeseries $\mathbf{A}$ and $\mathbf{B}$ is defined by
\begin{equation}\label{eq:mutinfdef}
	I^{AB} = H(\mathbf{A}) + H(\mathbf{B}) - H(\mathbf{A},\mathbf{B})
\end{equation}
where $H(\mathbf{A})$ is the entropy measured for timeseries $\mathbf{A}$ and $H(\mathbf{A},\mathbf{B})$ is the joint entropy, measured from a joint histogram. For a discrete timeseries, the Shannon entropy is defined by \cite{Shannon} 
\begin{equation}\label{eq:shannon}
	H = -\sum_i p_i \log_2 p_i
\end{equation}
where $p_i$ is again the probability of observing symbol $i$ and the summation is taken over all $i$.

There are two standard algorithms for computing the entropy $H$. The first, \cite{Fraser86}, discretizes the data using a hierarchy of partitions which become finer in regions of the joint histogram that contain more points. The second approach, \cite{Prichard95}, uses the second Renyi entropy \cite{Renyi70} which can be approximated by the logarithm of the correlation sum. Hence we can write the second Renyi mutual information as
\begin{equation}\label{eq:num33}
	I_2^{AB} = \log_2 C^{AB} - log_2 C^A - log_2 C^B
\end{equation}
where $C^{AB}$ is the joint correlation sum, which is the recurrence rate of the following type of cross recurrence plot
\begin{equation}\label{eq:crpdef}
    T_{ij}^{AB} = T_{ij}^A  T_{ij}^B
\end{equation}
This definition of a cross recurrence plot differs from the standard definition \cite{Zbilut98}, but has been recently proposed by Romano et al. \cite{Romano03} as a visualization of recurrent structure common to two timeseries. Thus we can obtain a standard mutual information estimate from three recurrence plots: $T_{ij}^A$, $T_{ij}^B$ and $T_{ij}^{AB}$.

The mutual information depends on the values of $C^A$ and $C^B$, which in turn are conditioned by $\epsilon^A$ and $\epsilon^B$, the threshold parameters used to produce the two auto recurrence plots. These two parameters must be chosen in some fashion, and this choice must be justified. One solution is to choose the thresholds such that the resulting auto recurrence plots have the same correlation sum. That is
\begin{equation}\label{eq:threshchoice}
	C^A(\epsilon^A) = C^B(\epsilon^B) = C_0
\end{equation}
This choice can be simplified by defining an unthresholded recurrence plot in terms of the measured correlation sum of the timeseries
\begin{equation}\label{eq:num36}
	U_{ij}^{A} = C_d^A(\|\mathbf{a}_i - \mathbf{a}_j\|)
\end{equation}
This recurrence plot has the property that if it is thresholded, then the resulting thresholded plot will have a recurrence rate (correlation sum) equal to the thresholding parameter. The corresponding unthresholded cross recurrence plot will now be given by
\begin{equation}\label{eq:num37}
	U_{ij}^{AB} = \max \{U_{ij}^A , U_{ij}^B \}
\end{equation}
since the definition of a thresholded recurrence plot uses the maximum norm Eq.~(\ref{eq:norm}). This allows us to write the joint correlation sum as a function of the elements of the joint recurrence plot
\begin{equation}
	C^{AB}(C_0) = \frac{2}{N(N-1)} \sum_{i=1}^N \sum_{j=i+1}^N \Theta(C_0-U_{ij}^{AB})
\end{equation}
Thus the joint correlation sum is equal to the recurrence rate of the unthresholded joint recurrence plot after it has been thresholded with a threshold parameter equal to $C_0$. Following Eq.~(\ref{eq:num33}) we then write the mutual information as
\begin{equation}\label{eq:num39}
    I^{AB}(C_0) = \log_2 C^{AB}(C_0) -2\log_2 C_0
\end{equation}

\begin{figure}
    \centering
    \includegraphics[width=0.6\textwidth]{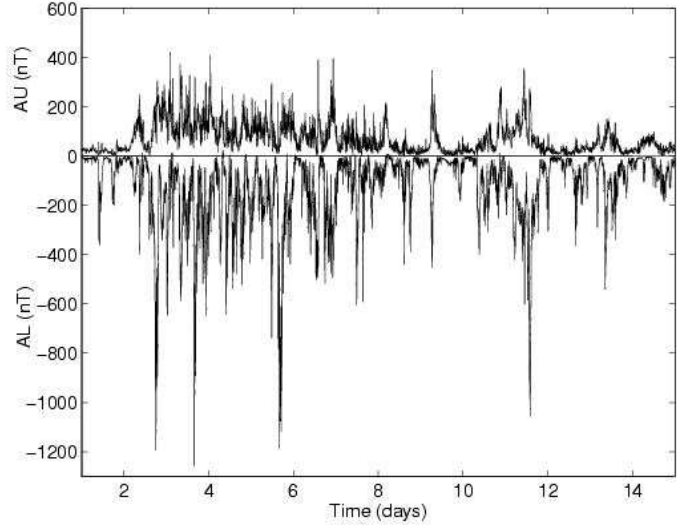}
  \caption{Days 1 to 14 of the AU and AL timeseries for the year 1995. AU, being the maximum reading from a network of magnetometer stations, is mostly positive, while AL is mostly negative.}
  \label{fig:AUAL}
\end{figure}

To demonstrate the quantitative practical use of this technique, we now apply it to the geomagnetic AU and AL timeseries for the year 1995. AU reflects the activity of eastward flowing polar currents, induced in the atmosphere by activity deeper in earth's magnetosphere. AL reflects the activity of westward currents, and is typically negative. Figure~\ref{fig:AUAL} shows these timeseries for the first two weeks of 1995. AU and AL typically come from opposite sides of the polar current system; they are therefore expected to share a certain amount of information due to large scale phenomena (storms) which are seen in both AU and AL, but to have differences due to smaller fluctuations arising from local phenomena. We use data for the entire year in order to get good statistics. Statistical noise acts to decrease the measured mutual information. The variance, due to noise, of mutual information measurements has been shown to scale with $1/N$ \cite{Roulston99}, where $N$ is the number of data points and here we have $N=5 \times 10^5$.

Within the AU and AL timeseries, three distinct classes of behavior are recognized phenomenologically: quiet time, storms and substorms. During quiet time, measurements of the order of a few nT to a few tens of nT are seen. The other extreme is seen during a magnetic storm, with measurements of hundreds of nT persisting for times of the order of several days. These events correlate strongly with features on the Sun facing the Earth \cite{McPherron95}, and thus tend to recur on a 27-28 day timescale (the synodic rotation period of the Sun). The intermediate event is a substorm \cite{Lyons96}, during which variations on the scale of tens to hundreds of nT persist for a few hours. Substorms are believed to result from the sudden release of stored energy built up in the magnetotail by the solar wind.

\begin{figure}
    \centering
    \includegraphics[width=0.45\textwidth]{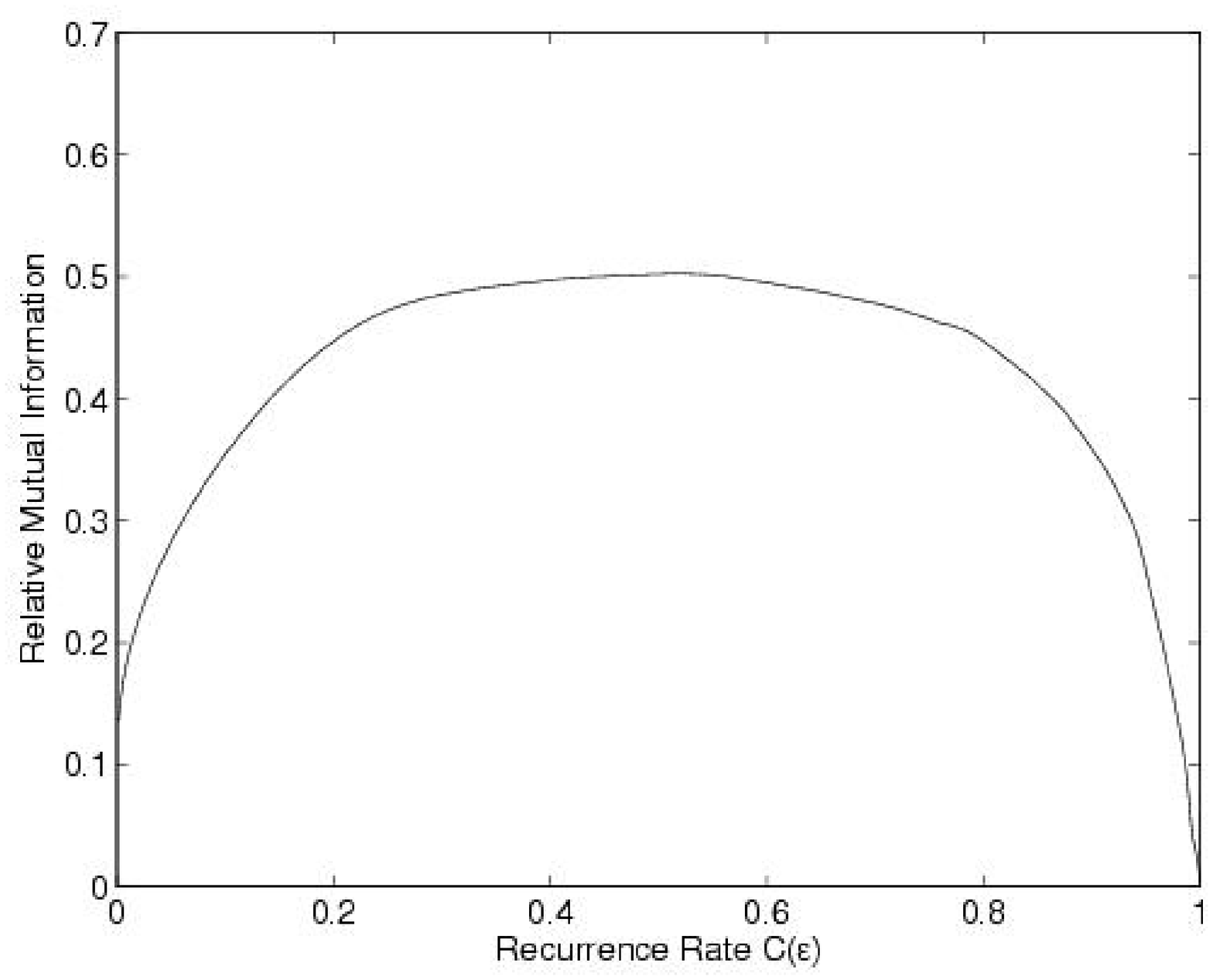}
    \includegraphics[width=0.45\textwidth]{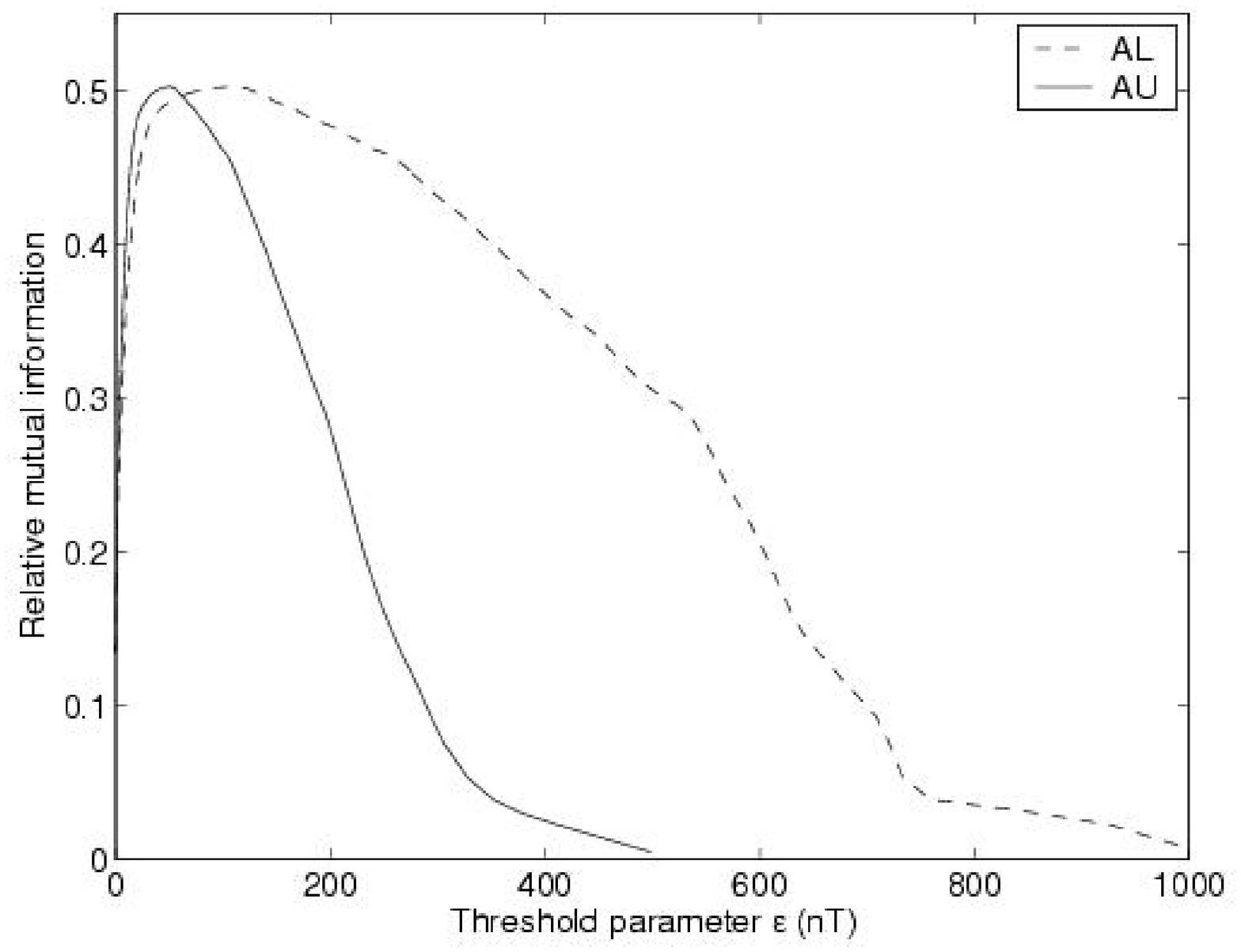}
  \caption{Mutual information $I$ for AU and AL geomagnetic timeseries, normalized to entropy of AU and AL separately. Left: as a function of correlation sum $C_0$, see Eq.~(\ref{eq:num39}); Right: as a function of the recurrence threshold parameter $\epsilon$ necessary to create the corresponding underlying thresholded recurrence plots for each measurement.}
  \label{fig:Itwiddle}
\end{figure}

Figure \ref{fig:Itwiddle} shows on the left the functional form of $I(C_0)$, the mutual information as a function of correlation sum of the underlying recurrence plots, obtained for the AU and AL timeseries. On the right are $I(\epsilon^A)$ and $I(\epsilon^B)$, the mutual information as a function of the underlying threshold parameters, constructed using Eq.~(\ref{eq:threshchoice}). Both figures are normalized to the entropy of AU and AL considered individually. The maximum fractional mutual information measured is $50\%$ and corresponds to an underlying correlation sum of $C_0=0.52$. To obtain this value of $C_0$, the two underlying thresholded recurrence plots require thresholds of $\epsilon$ = 49nT for AU and $\epsilon$ = 103nT for AL. On the right of Fig. \ref{fig:Itwiddle} is the functional form of $I(\epsilon)$, the mutual information as a function of the thresholds applied to the two underlying thresholded recurrence plots, again normalized to the entropy of AU and AL. The solid line shows the relative mutual information as a function of the threshold applied to the AU recurrence plot, while the dotted line shows the same for AL.

\begin{figure}
    \centering
    \includegraphics[width=0.6\textwidth]{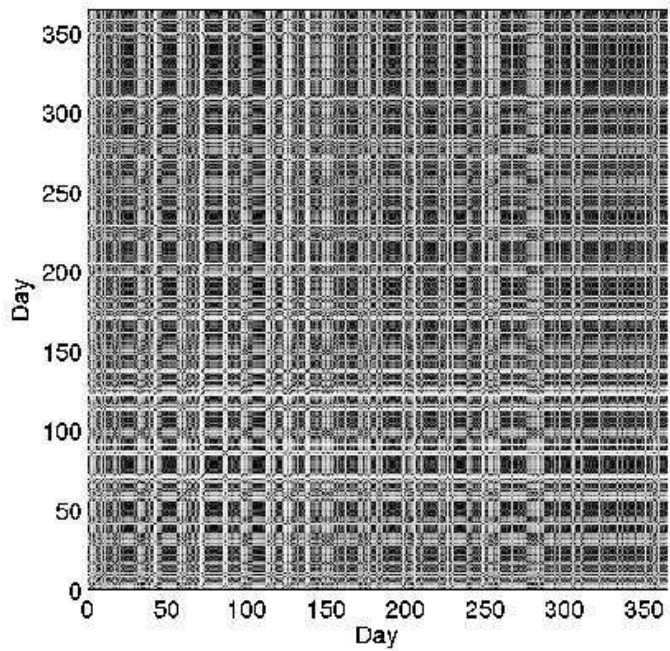}
  \caption{Unthresholded recurrence plot of geomagnetic AU timeseries.}
  \label{fig:AURP}
\end{figure}

\begin{figure}
    \centering
    \includegraphics[width=0.6\textwidth]{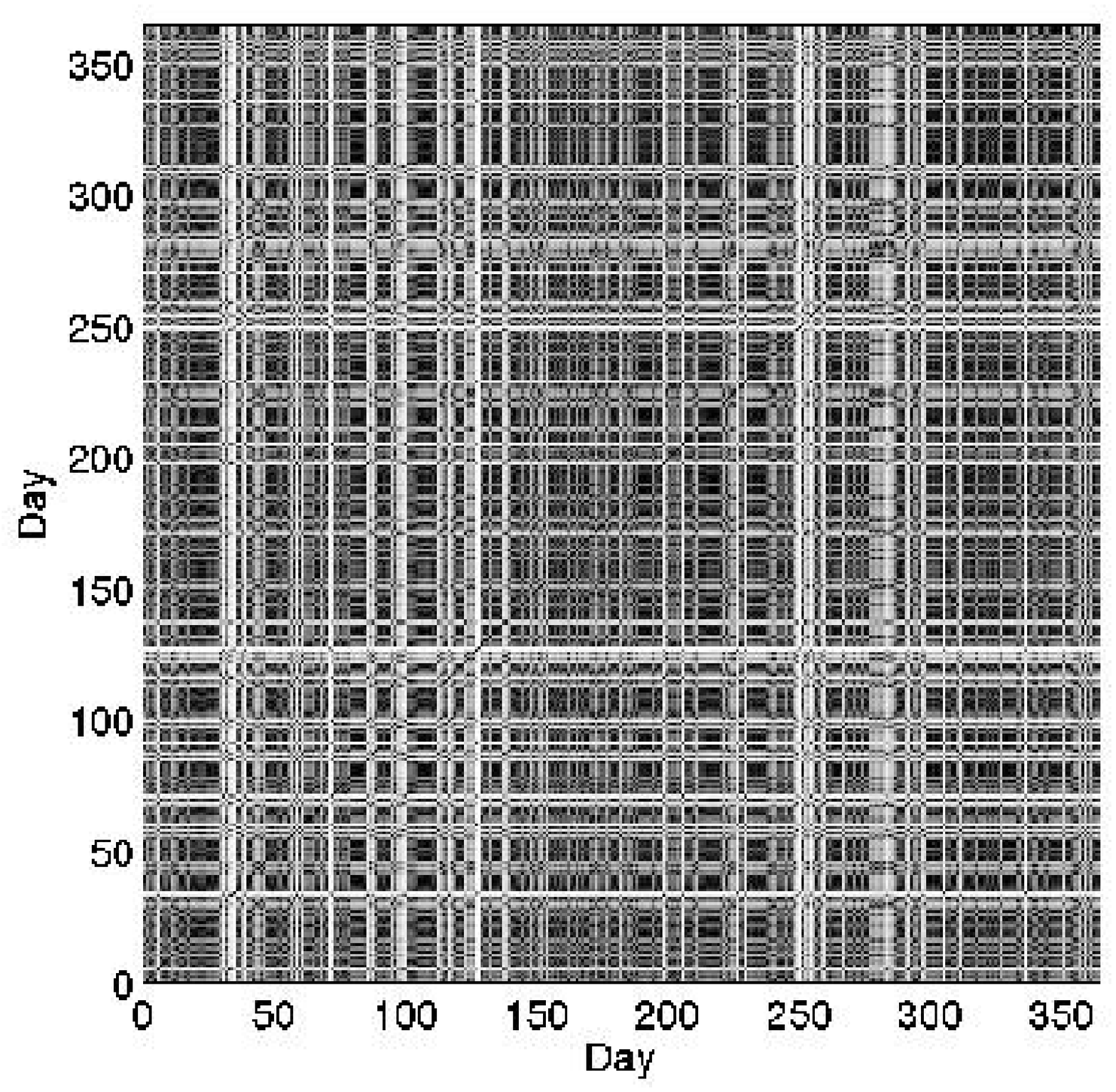}
  \caption{Unthresholded recurrence plot of geomagnetic AL timeseries.}
  \label{fig:ALRP}
\end{figure}

\begin{figure}
    \centering
    \includegraphics[width=0.6\textwidth]{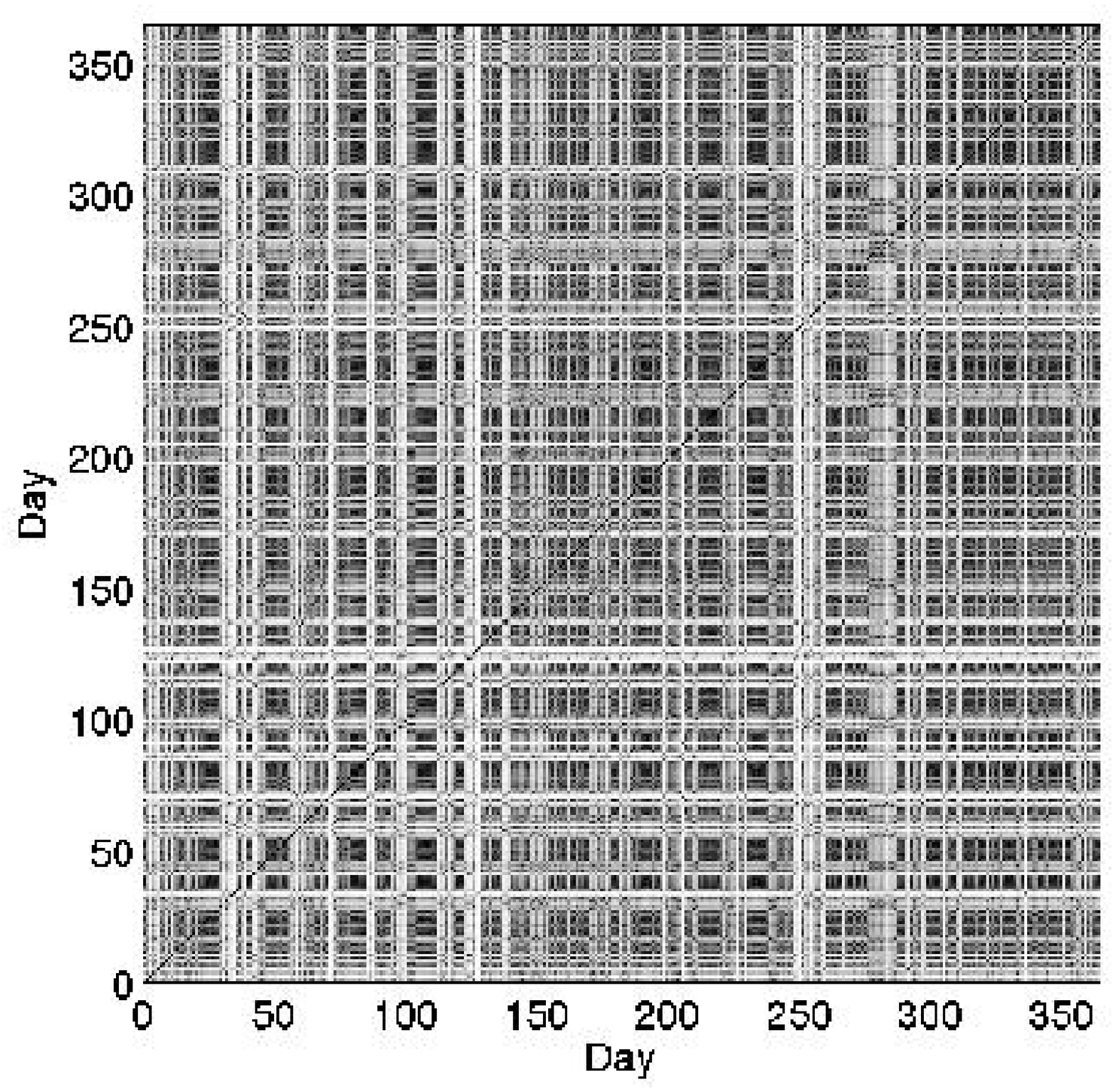}
  \caption{Unthresholded cross recurrence plot formed from those shown in Figs. \ref{fig:AURP} and \ref{fig:ALRP}.}
  \label{fig:AUALRP}
\end{figure}

Figures \ref{fig:AURP} and \ref{fig:ALRP} show unthresholded recurrence plots, as defined by Eq.~(\ref{eq:num36}) for the AU and AL geomagnetic timeseries respectively. The cross recurrence plot formed from these using Eq.~(\ref{eq:num37}) is shown in Fig. \ref{fig:AUALRP}. These plots show which positions in the original timeseries contribute the most to the mutual information -- in this case the dark areas on the cross recurrence plot correspond to the gaps between magnetic storms. We conclude that the mutual information being measured between AU and AL results from magnetic storms appearing in both timeseries.

\section{Conclusions}

Recurrence plots are extremely versatile: they analyse a stream of data by comparing segments of it to other segments taken at earlier and later times. The data stream itself is thus used as an analysis tool, without any assumptions about the nature of the process that produced it. There are many statistical measures associated with recurrence plots, some of which are unique to recurrence plot analysis. Here we have described two of the most common statistics, and have demonstrated that they are related to better known measures from nonlinear timeseries analysis. In the case of exponential scaling of the correlation sum with embedding dimension, the determinism and entropy of line length distribution have been shown to be determined by $K_2$. This explains the results of \cite{Iwanski98} and \cite{Zbilut00}.

We have also shown that all recurrence plots are contained within a single parent plot which contains all of the statistics of its children. It is not strictly necessary to construct recurrence plots for a variety of embedding parameters, because the key statistics that we have considered are all contained within this parent plot, and many of these are directly derivable from the distribution of diagonal line lengths. This demonstrates clearly the effect of embedding on recurrence plots.

A further result is that the mutual information between two timeseries can be obtained from their recurrence plots, and is related to counting the number of shared black dots. Similar comparisons of unthresholded recurrence plots yield the mutual information as a function of the threshold parameter $\epsilon$. This allows time-localized contributions to the mutual information to be assessed and quantified, as we have shown for the example of geomagnetic indices.

Comparisons between repeated patterns in signals from nonlinear systems are
particularly valuable when the systems in question are spatially extended
and evolve in a nonstationary fashion. Macroscopic plasmas, whether
naturally occuring or created in fusion experiments, often fall into this
category, which presents a substantial challenge to the techniques of
statistical and time series analysis; see, for example, the discussions in
recent studies of astrophysical\cite{GreenhA&A}, solar\cite{GreenhA&ALett},
and fusion\cite{GreenhPPCF} plasma observations. The successful application
of recurrence plots and the concepts of information theory to the
geomagnetic plasma timeseries studied in the present paper is, therefore,
encouraging.

{\bf Acknowledgements}
This work was funded in part by the United Kingdom Engineering and Physical Sciences Research Council, and by the Particle Physics and Astronomy Research Council.

\bibliographystyle{plain}

\begin{thebibliography}{10}

\bibitem{Maizel81}
J.~V. Maizel~Jr., R.~P. Lenk, Enhanced graphic matrix analysis of nucleic acid
  and protein sequences, Proceedings of the National Academy of Science USA 78
  (1981) 7665--7669.

\bibitem{Eckmann87}
J.~P. Eckmann, S.~O. Kamphorst, D.~Ruelle, Recurrence plots of dynamical
  systems, Europhys. Lett. 4 (1987) 973--977.

\bibitem{Webber94}
C.~L. Webber~Jr., J.~P. Zbilut, Dynamical assessment of physiological systems
  and states using recurrence plot strategies, J. Appl. Physiol. 76 (1994)
  965--973.

\bibitem{Faure98}
P.~Faure, H.~Korn, A new method to estimate the {K}olmogorov entropy from
  recurrence plots: its application to neuronal signals, Physica D 122 (1998)
  265--279.

\bibitem{Foote01}
J.~Foote, M.~Cooper, Visualizing musical structure and rhythm via
  self-similarity, in: Proceedings International Conference on Computer Music,
  Havana, Cuba, 2001.

\bibitem{Marwan02b}
N.~Marwan, J.~Kurths, Nonlinear analysis of bivariate data with cross
  recurrence plots, Physics Letters A 302 (2002) 299--307.

\bibitem{Marwan02c}
N.~Marwan, N.~Wessel, U.~Meyerfeldt, A.~Schirdewan, J.~Kurths,
  Recurrence-plot-based measures of complexity and their application to
  heart-rate-variability data, Physical Review E 66 (2002) 026702.

\bibitem{Bernstein91}
M.~Bernstein, J.~D. Bolter, M.~Joyce, E.~Mylonas, Architectures for volatile
  hypertext, in: Proceedings of the Third Annual ACM Conference on Hypertext,
  San Antonio, Texas, United States, 1991, pp. 243--260.

\bibitem{Cutler00}
R.~Cutler, L.~Davis, Robust periodic motion and motion symmetry detection, in:
  Proc. Conference on Computer Vision and Pattern Recognition, South Carolina,
  USA, 2000.

\bibitem{Church93}
K.~W. Church, J.~I. Helfman, Dotplot: a program for exploring self-similarity
  in millions of lines of text and code, J. American Statistical Association 2
  (1993) 153--174.

\bibitem{Takens81}
F.~Takens, Detecting Strange Attractors in Turbulence, Vol. 898 of Lecture
  Notes in Math, Springer, New York, 1981.

\bibitem{Fraser86}
A.~M. Fraser, H.~L. Swinney, Independent coordinates for strange attractors
  from mutual information, Phys. Rev. A 33 (1986) 1134--1140.

\bibitem{Iwanski98}
J.~S. Iwanski, E.~Bradley, Recurrence plots of experimental data: To embed or
  not to embed?, Chaos 8 (1998) 861--871.

\bibitem{KantzSchreiber}
H.~Kantz, T.~Schreiber, Nonlinear time series analysis, Cambridge University
  Press, 1997.

\bibitem{Gao00}
J.~Gao, H.~Cai, On the structures and quantification of recurrence plots,
  Physics Letters A 270 (2000) 75--87.

\bibitem{AE}
T.~N. Davis, M.~Sugiura, Auroral electrojet activity index {AE} and its
  universal time variations, Journal of Geophysical Research 71 (1966)
  785--801.

\bibitem{Hnat02}
B. Hnat, S. C. Chapman, G. Rowlands, N. W. Watkins, M. P. Freeman, Scaling in solar wind epsilon and the AE, AL and AU indices as seen by WIND, Geophys. Res. Lett. 10 (2002) 1029.
  
\bibitem{Casdagli97}
M.~C. Casdagli, Recurrence plots revisited, Physica D 108 (1997) 12--44.

\bibitem{Zbilut00}
J.~P. Zbilut, A.~Giuliani, C.~L. Webber~Jr., Recurrence quantification analysis
  as an emperical test to distinguish relatively short deterministic versus
  random number series, Physics Letters A 267 (2000) 174--178.

\bibitem{randomorg}
M.~Haahr, http://www.random.org.

\bibitem{Shannon}
C.~E. Shannon, W.~Weaver, The mathematical theory of communication, University
  of Illinois Press, 1949.

\bibitem{Prichard95}
D.~Prichard, J.~Theiler, Generalized redundancies for time series analysis,
  Physica D 84 (1995) 476--493.

\bibitem{Renyi70}
A.~Renyi, Probability theory, North Holland, Amsterdam, 1970.

\bibitem{Zbilut98}
J.~P. Zbilut, A.~Giuliani, C.~L. Webber~Jr., Detecting deterministic signals in
  exceptionally noisy environments using cross-recurrence quantification,
  Physics Letters A 246 (1998) 122--128.

\bibitem{Romano03}
M.~Romano, M.~Thiel, J.~Kurths, A new definition of cross recurrence plots,
  submitted to Physics Letters A (2003).

\bibitem{Roulston99}
M.~S. Roulston, Estimating the errors on measured entropy and mutual
  information, Physica D 125 (1999) 285--294.

\bibitem{McPherron95}
M.~G. Kivelson, C.~T. Russell (Eds.), Introduction to Space Physics, Cambridge
  University Press, 1995, Ch.~13.

\bibitem{Lyons96}
L.~R. Lyons, Substorms: Fundamental observational features, distinction from
  other disturbances, and external triggering, Journal of Geophysical Research
  101 (1996) 13011--13025.

\bibitem{GreenhA&A}
J.~Greenhough, S.~C. Chapman, S.~Chaty, R.~O. Dendy, G.~Rowlands,
  Characterising anomalous transport in accretion disks from {X}-ray
  observations, Astronomy and Astrophysics 385 (2002) 693--700.

\bibitem{GreenhA&ALett}
J.~Greenhough, S.~C. Chapman, R.~O. Dendy, V.~M. Nakariakov, G.~Rowlands,
  Statistical characterisation of full-disk {EUV/XUV} solar irradiance and
  correlation with solar activity, Astronomy and Astrophysics 409 (2003)
  L17--L20.

\bibitem{GreenhPPCF}
J.~Greenhough, S.~C. Chapman, R.~O. Dendy, D.~J. Ward, Probability distribution
  functions for {ELM} bursts in a series of {JET} tokamak discharges, Plasma
  Physics and Controlled Fusion 45 (2003) 747--758.

\end{thebibliography}

\end{document}